\documentclass[preprint]{revtex4}
\usepackage{bm}
\input{epsf}
\begin{document}
\baselineskip=28pt
\title{Probing Internal Stress and Crystallinity in Wet Foam \emph{via} Raman Spectroscopy}
\author{T. K. Barik, P. Bandyopadhyay and A. Roy}
\email{anushree@phy.iitkgp.ernet.in} \affiliation{Department of
Physics, Indian Institute of Technology Kharagpur 721 302, India}
\begin{abstract}
In this article, we correlate the internal stress and the
characteristics of a vibrational mode in wet foam. Using
microscope images, we estimate the average size of the bubbles in
wet foam, at specific time intervals, over a duration of twenty
four hours. Raman spectra are also recorded at the same time
intervals, over the same time frame. We show that the internal
stress, originated from the macroscopic structural change of foam
with ageing, can be related to the observed Raman shift of the low
frequency methylene rocking mode of the constituent surfactant
molecules in foam. In this report we also show the capability of
the Raman spectroscopy to reveal the crystallinity in foamy
materials, when studied for a longer period of time.

\end{abstract}
\maketitle
\def\d{{\mathrm{d}}}

\section{Introduction}

Soft foam, a cellular fluid, is a two phase system. It consists of
a collection of gas bubbles surrounded by thin liquid films. With
time, the spherical bubbles in fresh foam take the form of
polyhedra while minimizing the energy of the system. It can
coarsen by the diffusion of gas from smaller bubbles to larger
bubbles \cite{Weaire}.
Furthermore, the liquid between the bubbles can drain out along
the liquid channels (Plateau borders) in response to gravity. The
adjacent bubbles coalesce if the liquid film becomes too thin.

Soft foam exhibits interesting elastic properties. Under low
applied shear stress, foam behaves as an elastic solid. With an
increase in stress it becomes progressively plastic; beyond a
certain yield stress, the foam flows along with topological
changes. Such characteristics of foamy structure strongly depends
on bubble size and wetness \cite{Fortes}. Both two- and three
dimensional foam can be accurately simulated using various models
\cite{Weaire2}. The computer simulation results suggest that, in
the low compression limit, there exists a correlation between the
shear modulus and gas/liquid fraction in the tightly packed gas
bubbles \cite{Bolton:1991a,
Feng:1985,Hutzler:1995,Princen:1986,Durian:1995,Durian:1997}.
These models, therefore, reveal a connection between the complex
macroscopic rheological behavior of foam and its underlying
microscopic structure.

The complex foam structure is composed of extended polyatomic
organic molecules (surfactants) and water. Raman spectroscopy is a
powerful noninvasive tool to probe the molecular structure and
dynamics of a system.
In wet foam, Raman scattering is caused by deformation/stretching
of different vibrational bonds of constituent molecules. Thus, if
it is assumed that macroscopic and microscopic behavior of wet
foam can be related, one expects that the analysis of Raman line
profiles, which reveal molecular behavior, can be used to probe
the elastic properties of wet foam. The stress induced shift in
the Raman lines has been reported for other soft matter, like soft
polymers and polymer based fiber structures \cite{Nickolov,
macromolecule}. For example, it has been shown by Ward and Young,
via polarized NIR Raman measurements, that Raman bands of
thermotropic aromatic copolyesters exhibit linear shifts towards
lower wavenumbers with stress and strain following the tensile
deformation \cite{Ward1}. The correlation between the architecture
of the network and elastic properties of their building blocks
leads to interesting mechanical properties of such systems
\cite{Heussinger}.
 The main difficulty in
using Raman spectroscopy to probe wet foam arises due to multiple
scattering of light within the bubbles. The broad and strong
background due to the scattered light masks the Raman signal from
the foamy structure by a large extent. Thus, in the literature, we
do not find too many articles on Raman studies of wet foam. The
most significant one is by Goutev and Nickolov \cite{Goutev:1996},
where the authors have studied
 the molecular structure of Gillette foam, to some
extent, using Raman spectroscopy. Recently, we have reported our
results on Raman measurements in Gillette foam, where we have
analyzed the evolution of the O-H vibrational bond of water
molecules with ageing \cite{Animesh}.

In this article, we probe, using Raman scattering, how the
macroscopic changes in structure of bubbles in wet foam is
manifested in changes in the molecular structure of the surfactant
molecules. Section II explains the reasons for the choice of our
sample and also describes other experimental details. The
structural changes in foam with time have been described in
Section III. In Section IV, we have shown the effect of structural
change in wet foam on the molecular vibrational modes of Gillette
foam, as obtained from Raman measurements. We correlate the change
in molecular vibrations with the evolution of internal stress and
crystalline structure of foam. Finally, in Section V, we summarize
our results with a few remarks.

\section{Experiment}

\begin{figure}[htbp]
\centerline{\epsfxsize=3in\epsffile{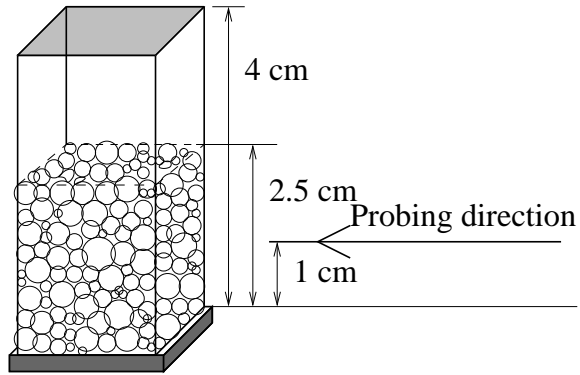}} \caption{The
schematic of the foam container and the probing direction.}
\label{fig1}
\end{figure}

Though complex in composition, Gillette shaving foam is often used
for studying optical properties of wet foam. It is reproducible
and stable over the duration required for optical measurements.
For Raman measurements this commercial foam offers an extra
advantage in the following sense. When laser light is incident on
foam, it undergoes multiple scattering. In order to obtain the
optimum Raman signal, the mean free path, $l^\star$,[ $\cong$ 3.5
$\times$ average diameter of the bubbles ($d$)] of light within
the foam should be comparable with the slit width of the
spectrometer collecting the scattered light \cite{Goutev:1996}.
The mean diameter of bubbles in fresh Gillette shaving foam is
close to 50 $\mu$m and the maximum diameter, which we have
studied, is $\sim$ 150 $\mu$m, which is comparable with the
slit-width of our spectrometer ($\sim$ 100 $\mu$m).

In this commercial foam, the basic ingredients [triethanolamine
stearate with small amount ($<1$\%) of sodium lauryl sulphate and
polyethylene glycol lauryl ether and emulsified liquid hydrocarbon
gases] are kept in an aqueous solution under high pressure. The
foam is produced after expansion of the above mixture in the
aqueous solution. The experiments have been carried out by taking
the foam from the can in a closed rectangular quartz cell of
dimension 1cm $\times$ 1cm $\times$ 4cm [Fig. \ref{fig1}]. In the
beginning, the material fills $\sim$ 2.5 cm of the whole cell
volume from the bottom. The cell is then sealed with paraffin
films to prevent direct evaporation. Raman measurements and
imaging experiments have been carried out at a distance 1 cm from
the bottom of the cell [Point P in Fig. \ref{fig1}]. To observe
the effect of ageing, the measurements have been continued at the
same point throughout the duration of the experiment. The
experiments have been repeated for two other heights on the column
of foam to check the reproducibility of the results.

The change in the structure of foam with aging, is probed through
optical images using a Metzer Biomedical microscope (model:
MEGA-6021). These images are analyzed using Image-J
image-processing software. Simultaneously, Raman spectra are
recorded in a back-scattering geometry using TRIAX550 single
monochromator equipped with a notch filter and CCD as a detector.
An argon ion laser of wavelength 488 nm and of power 30 mW on the
sample has been used as an excitation source. For more details of
the technique and spectrometer, see \cite{Roy:AJP}. As the spot
size of the laser, used as an excitation source in Raman
measurements, is $\sim$ 0.5 mm (much more than the size of the
channels between bubbles or Plateau border region); at present, it
is not possible for us to distinguish, whether the measured Raman
signals are confined only to the Plateau border region or within
the thin film between two bubbles or in both. All measurements and
analysis of data have been carried out for three different sets of
experiments to check the reproducibility of the results. Raman
spectra are fitted with Lorentzian line shape keeping peak
position, width and intensity as fitting parameters in order to
estimate them properly, for each observed feature.

\begin{figure}
\centerline{\epsfxsize=6in\epsffile{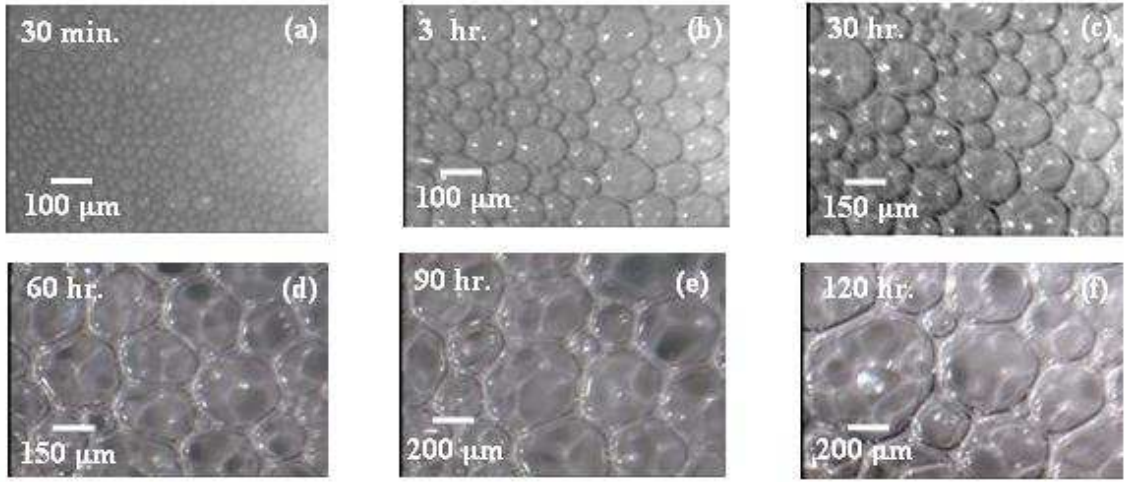}}
\caption{Microscopic images of liquid foam at different time
scale: the corresponding magnification of the images are shown in
each figure.} \label{fig2}
\end{figure}

\section{Shape and Structure}


\begin{figure}
\centerline{\epsfxsize=4in\epsffile{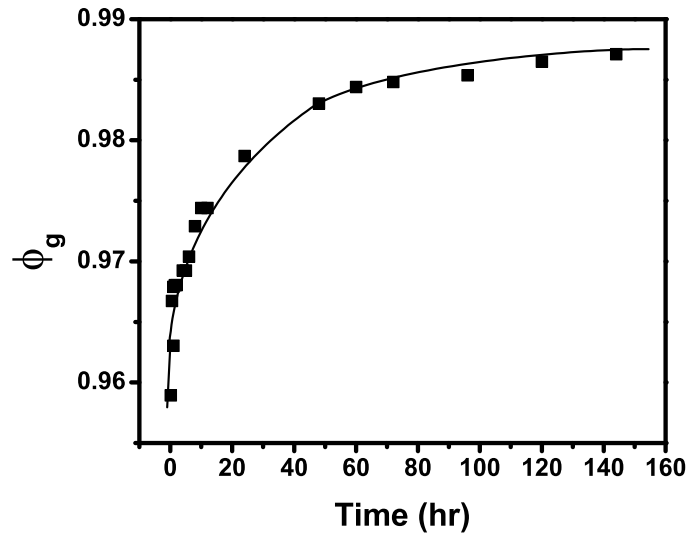}}
\caption{Variation in gas fraction with aging. The solid line is a
guide to the eye.} \label{fig3}
\end{figure}

As mentioned earlier, aging effects on foam structure can be seen
through microscopy. More precisely, we recorded the images of foam
at an interval of 15 minutes in the beginning (till 12 hrs) and
less frequently at the end, over 7 days. Fig.\ref{fig2}(a) to (f)
show few characteristic microscopic images at different stages of
aging. Initially, the bubbles are spherical, separated by Plateau
borders touching the quartz cell wall. Subsequent snapshots show
that the bubbles take on a polyhedral structure due to coarsening
of the bubbles and liquid drainage.

The three-dimensional (3D) gas fraction, ($\phi_{g}$), is the
commonly used parameter to describe a foam.  $\phi_{g}$ can be
estimated from the following relation
\begin{equation}
\phi_{g} = 1-\frac{\rho_{1}}{\rho_{2}},
\end{equation}
Here,  $\rho_{1}$ and $\rho_{2}$ are the mass densities of the
foam and the liquid phase ($\rho_{2}$=0.997 gm/cc \cite{density}).
Foam mass density was obtained from the measured weight of the
foam. The variation in 3D-gas fraction with time is shown by
filled square in Fig. \ref{fig3}(b). The solid line is the guide
to the eye. It is to be noted that the gas fraction varies for the
foam expelled from different portions of the can
\cite{Goutev:1996}.

\section{Molecular Vibrational bands - Results and Discussion}

\subsection{Low frequency methylene rocking mode}
\begin{figure}
\centerline{\epsfxsize=4in\epsffile{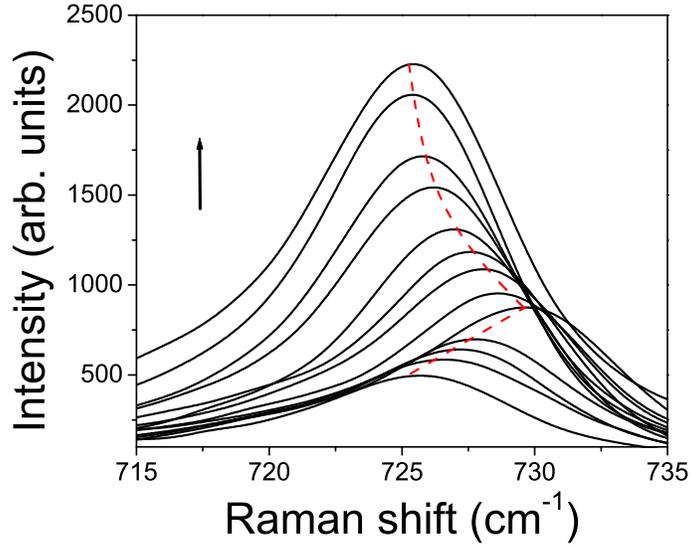}} \caption{The
variation in the methylene rocking modes due to aging of Gillette
foam. The direction of the arrow is along the increase in average
diameter of bubbles (43,75,93,117,135,144,157,182,222, 283,318,
400 and 440 ${\mu}$m). The red dashed curve is a guide to the eye
to show the non-monotonic variation in Raman shift.} \label{fig4}
\end{figure}

In the Raman spectrum of Gillette foam, the time evolution of the
out of plane methylene rocking modes, at around 725 cm$^{-1}$
\cite{Snyder:1967}, is shown in Fig. \ref{fig4}. The intensity is
scaled to show the variation in spectral frequency with the ageing
of foam. Here, we have shown a few characteristic spectra; though
we recorded the Raman spectrum over the same period and at similar
intervals, as was done in the imaging experiment. This Raman line
corresponds to the trans- conformation of the methylene chain. The
direction of the arrow is along the increase in time (the average
diameter of the bubbles increases during the ageing process).
Interestingly, the Raman shift of the peak at around 725 cm$^{-1}$
exhibits a non-monotonous behavior with ageing (see the dotted red
line).

At this point we draw an analogy between the behavior of a solid
and a foam network. For solids, the magnitude of the shift in
Raman wavenumber for the modes of quantized lattice vibration can
be related to the component of the stress tensor of the system
along different directions via a constant, which has the unit of
frequency change per unit stress \cite{Cardona}. A compressive
stress results in a higher wavenumber  shift in the Raman line,
whereas, a tensile stress shifts the Raman line in the opposite
direction. To the best of our knowledge, any such study on a foamy
structure is not available in the literature. Nevertheless, it is
to be noted that soft matter exhibits behavior, which is, at
times, closer to that of a solid. The surfactant molecules are
fastened together in a group rather rigidly and are constrained at
the mesoscopic scale to behave more like a solid. The interaction
of the constituent molecules within these groups determine the
macroscopic behavior of foam. On the other hand, these molecular
groups do move like the molecules in a simple fluid \cite{Aubert}.
In view of these facts,
we attempt to explain the non-monotonous change in Raman shift, as
shown in Fig. \ref{fig4}, by assuming a correlation between the
Raman shift of the vibrational mode and the internal stress in wet
foam, as done in the case of solids.

\begin{figure}
\centerline{\epsfxsize=4in\epsffile{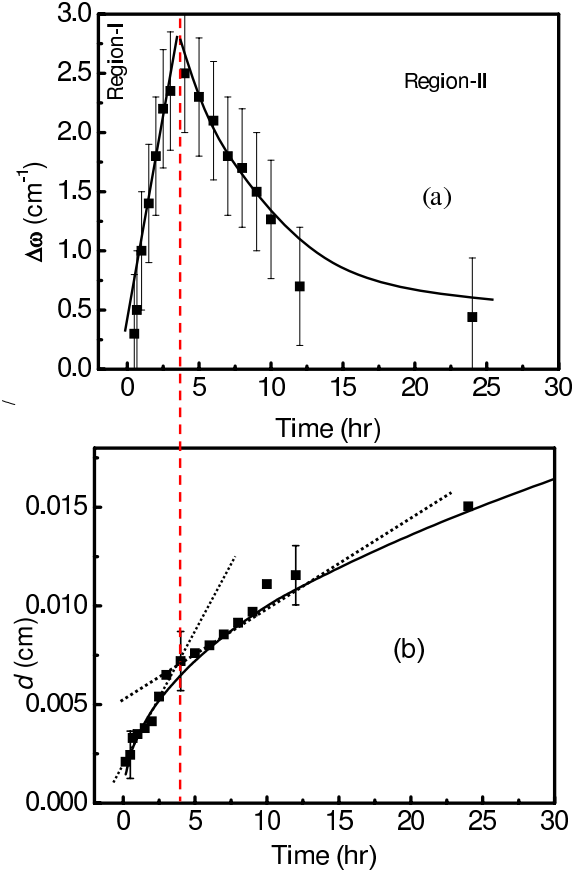}} \caption{(a) A
plot of $\Delta{\omega}$ vs $t$ for methylene rocking mode as
observed from Raman measurements (filled squares) as obtained from
Fig. 4. The solid line is a guide to the eye. (b)Variation in
average bubble diameter with aging.}
\label{fig5}
\end{figure}

We define the difference between the Raman frequency ($\omega_t$)
of the above mode from the surfactant of foam at time $t$, and the
Raman frequency of the same ($\omega_0$) in fresh Gillette foam as
$\Delta{\omega}$ ($\Delta{\omega} = \omega_{t}-\omega_0$). The
variation of $|\Delta{\omega}|$ with $t$, is shown in Fig.
\ref{fig5}(a) [filled squares]. The magnitude of the difference
($\Delta \omega$) increases with increase in $t$ till
$t=t^{\prime}$ $\sim$ 4 hr (Region I, the left hand side of the
dashed-red line, in Fig. 5), indicating a compressive stress in
the liquid film to be the origin of the shift. Then it decreases
with a further increase in time (Region II, the right hand side of
the dashed-red line, in Fig. 5), indicating a gradual release of
the stress.

To explain the above variation of stress with time in wet foam, we
take a re-look at the variation in average bubble diameter $d$
with time (a few characteristic image frames are shown in Fig. 2).
Assuming the bubbles to be nearly spherical at all stages, we
measured the average diameter of 200 bubbles for each time frame.
In Fig. \ref{fig5}(b) the variation in measured $d$ with time is
shown by the filled squares.  The time scale of evolution of the
cellular pattern can be taken to be inversely proportional to the
length scale. Hence, the average diameter of the bubbles follows
the scaling behavior \cite{Weaire3,Durian:1991},
\begin{equation}
d \propto (t-t^{\prime})^{1/2}.
\end{equation}
Here, $t^{\prime}$ is a constant. In Fig. \ref{fig5}(b) the best
fitted line to the data points with Eqn. 2, is shown by the solid
line. In a strict sense, Eqn. 2 is valid for dry foam
\cite{stine:1990}. This explains the slight deviation of the
fitted line from the experimental data in Fig. \ref{fig5}(b). Here
it is interesting to note that the smooth variation of $d$ with
$t$ in Fig. 5 (b) can be decomposed into two parts with two
distinct slopes (shown by dotted black lines). Again we find that
in Region I, ie. till time $t$= 4 hr. (same time during which the
compressive stress on the molecules reaches maximum in Fig.
\ref{fig5}(a)) the increase in $d$ with $t$ is relatively sharp,
then it gradually slows down.

When the diameter of a bubble in fresh foam fluctuates to one
which is infinitesimally larger in size, the diameter of the
neighboring bubble shrinks by  the same amount. Due to the larger
internal pressure in the smaller bubble, the air flows from it to
the larger bubble. As a result, the size of the smaller bubble
decreases further in size. This phenomenon causes an increase in
average bubble diameter in foam. The relatively rapid increase in
diameter of the bubble in Region I in Fig. \ref{fig5} (b)
compresses the liquid film between bubbles, resulting in a
compressive stress in the film (a `jamming' effect), shown in
Region I of Fig. \ref{fig5} (a). At this point we refer to the
pioneering work by Friberg and Langlois \cite{Friberg}, where it
has been shown that two phases can exist in wet foam. Though the
liquid phase in foam is known to be isotropic at the initial
stage, with ageing, the formation of lamellar phase results in the
crystal-like structure in foam \cite{Goutev:1996}. The small
lamellae gets parallel to the interface and gradually forms a
large bilayer structure. The interface of gas-liquid has a
noticeable effect on this lamellar phase, more than what is
observed for the isotropic phase \cite{Tiddy}. For a spherical
bubble in a liquid, the pressure difference between the gas-liquid
interface normally confirms the Laplace-Young law. The Laplace
pressure acting outward on the surface is given by
$\Delta{p}=\frac{4\gamma}{R}$, with $\gamma$ being the surface
tension between gas-liquid interface and $R$ being the local
radius of curvature of the surface. Due to the formation of
lamellar phase in foam at a later time, the pressure acting in the
film between bubbles is gradually released with an increase in
$R$. Also note that at a later time (Region II in Fig. 5), the
growth rate is also less. In addition, we need to keep in mind
that from a wet foam liquid drains out with time. All three
effects, effectively, cause a release in pressure in the film.
Fig. \ref{fig5}(a) and (b) possibly indicate two coupled and
competitive dominating phenomena (coarsening and drainage) in the
ageing process of the foam.


In the above, we have tried  to  explain the variation of the
Raman frequency of one of the strongest molecular vibrational
modes of Gillette foam by taking into account the internal stress
in the system.
We would like to mention that, such an effect of aging on Raman
shift has not been observed for other vibrational modes in wet
foam \cite{Goutev:1996}.  A possible explanation can be that the
other features corresponding to stretching molecular vibrations,
are, usually, convoluted with many other modes and therefore, this
effect is masked. Moreover, the effect of internal shear (which
eventually cancels out when we take ensemble average) in foam is
expected to be more for angular deformation of the local bonds
rather than bond stretching.

\subsection{Characteristics of other C-H vibrations}

\begin{figure}
\centerline{\epsfxsize=3.2in\epsffile{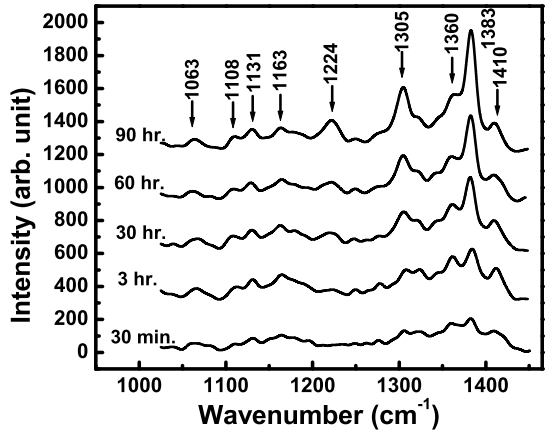}} \caption{The
change in Raman spectrum of Gillette foam with aging in the region
between 1000 and 1450 cm$^{-1}$. } \label{fig7}
\end{figure}

To illustrate the time evolution of the molecular structure of
Gillette foam, few characteristic Raman spectra over the range
between 1000 and 1450 cm$^{-1}$ are shown in Fig. \ref{fig7}. The
experiments have been carried out for 7 days. The main features
are indicated by arrows. Four types of vibrations are observed,
namely stretching and deformation of C-H and C-C bonds. The peaks
at 1063 and 1131 cm$^{-1}$ are the in-phase and out-of-phase C-C
rocking vibrational modes for the functional group
$^{^{\backslash}}_{_{/}}$C(CH$_{3})_2$. These features indicate
trans intra-molecular conformation. Here we would like to point
out that for gauche conformation, the Raman lines are expected to
appear between 1085 and 1095 cm$^{-1}$, which are absent in  Fig.
\ref{fig7}. In addition, here we mention, especially, the
following features.  The C-C skeletal vibration in -C(CH$_{3})_3$
functional group and C-H symmetric vibration of the
 -CH$_3$, which appear at 1224 cm$^{-1}$ and
1383 cm$^{-1}$. One expects two overlapping bands, one at 1368 and
the other at 1352 cm$^{-1}$, due to C-H deformation vibrations in
CH. These two features merge and appear as one peak at 1360
cm$^{-1}$. The C-H deformation vibration in -(CH$_{2})_{n}$-
appears at around 1305 cm$^{-1}$, the intensity of this feature in
expected to increase with $n$. All the above features along with
others, shown by arrows in Fig. \ref{fig7}, and their assignments
are tabulated in Table I.

\begin{table}[htbp]
\caption{Assignment of Raman vibrational bands between 1000 to
1450 cm$^{-1}$ for Gillette foam}
\begin{tabular}{|c|c|c|} \hline
Assignment  &  Wavenumber (cm$^{-1}$) & Ref. \\ \hline
C-C rocking mode in $^{^{\backslash}}_{_{/}}$C(CH$_{3})_2$ & 1063 \& 1131 & \cite{Goutev:1996}\\
C-C stretching mode in  $^{^{\backslash}}_{_{/}}$C(CH$_{3})_2$ & 1163&\cite{Socrates}\\
C-C stretching in phospholipid bilayer & 1108 & \cite{Goutev:1996}\\
C-C skeletal vibration in -C(CH$_{3})_3$ & 1224& \cite{Socrates}\\
C-H deformation vibration in -(CH$_{2})_{n}$& 1305 &  \cite{Snyder:1967}\\
C-H deformation vibrations in CH & 1360 & \cite{Socrates}\\
C-H symmetric vibration in -CH$_3$ & 1383 &\cite{Socrates}\\
acyl chain of polyethylene& 1410 & \cite{Yellin:1977}\\

\hline
\end{tabular}
\end{table}

\begin{figure}
\centerline{\epsfxsize=3.2in\epsffile{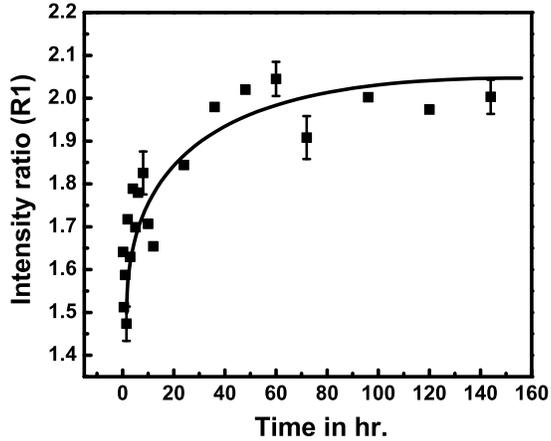}} \caption{The
variation of intensity ratio (R1) of the Raman lines at 1305 and
1383 cm$^{-1}$ with aging of foam.}\label{fig8}
\end{figure}

From Fig. \ref{fig7}  it is clear that except the intensities of
the features at 1383,1305, and 1224 cm$^{-1}$, the intensities of
the other peaks remain nearly constant with aging. The increase in
intensities of the Raman lines at 1224 and 1305 cm$^{-1}$ indicate
formation of longer polymeric chains, (CH$_{3})_{3}$ and
(CH$_{2})_{n}$, with the aging of foam. Vibrational spectra of the
C-H stretching region are complex due to Fermi-resonance
interactions  between the symmetric methylene C-H stretching mode
and the overtones of the CH$_2$ rocking modes. Unlike the C-H
stretching vibrations, the CH$_2$ rocking modes participate
significantly in intermolecular coupling and are thus influenced
by the lateral packing order. Due to the Fermi resonance
interaction of CH$_2$ rocking with symmetric C-H stretching, the
latter indirectly gets affected by the lateral packing order. On
the other hand, the asymmetric C-H stretching mode is forbidden by
symmetry for Fermi-resonance interactions and is thus insensitive
to the lateral ordering. Thus, one can expect that the asymmetric
vibrational mode at 1305 cm$^{-1}$ should depend only on chain
conformation, while the symmetric vibrational mode at 1382
cm$^{-1}$ should be modified by the lateral packing order.
Consequently, the ratio (R1) of intensity of the feature at 1305
cm$^{-1}$ to that at 1383 cm$^{-1}$ can be used as a parameter
which is sensitive to both the lateral packing  and to
conformational order within the chains. The variation of the ratio
R1 with $t$ is shown in Fig. \ref{fig8}. The value of this ratio
is expected to be 0.7 for completely melted hydrocarbon chains,
1.5 for vibrationally decoupled all-trans chains, and 2.2 for a
highly ordered crystalline lattice \cite{Dierker:1987}. From Fig.
\ref{fig8} it is clear that the surfactant molecules undergoes a
gradual structural change with time.  After $\sim$ 40 mins the
value of R1 reaches the value $\sim$ 1.6, indicating that the
constituent molecules are in trans conformation, however gradually
they organize into an ordered multilayer or crystal structure
(with R1 = 2.1).

\section{Summary}
As far as we know, this is the first report where Raman
spectroscopy has been used to determine the internal stress in wet
foam. In other words, here we have shown that Raman spectroscopy
has the potential to extract information about internal stress in
the system. We have related the observed shift in the low
frequency Raman peak position of the methylene rocking mode with
the variation in internal stress in the system.

The composition of commercial shaving foams is quite complex and
its physico-chemical properties are ill defined. Though, our
method can be used as a quick and noninvasive tool to measure the
strain and hence,
 the stability,  of a commercial foam;
it is worth to check the above claim for  simple foamy material
with well controlled composition, specially made in a laboratory.
Further, experiments on known surfactants will also indicate if
the observed behavior of the wet foam originates from the
characteristics of the surfactant itself or from its foamy
structure.

\section{Acknowledgements}
The authors thank A.K. Sood, IISc, Bangalore; A.K. Sharma, IIT,
Kanpur; R. Bandyopadhyay, RRI, Bangalore, for their valuable
comments.They are also indebted to T. Pathak, IIT Kharagpur, for
his comments on the synthesis of surfactants.

\end{document}